# Visualizing Squircular Implicit Surfaces

Chamberlain Fong
chamberlain@alum.berkeley.edu

**Abstract** – The squircle is an intermediate shape between the square and the circle. In this paper, we examine and discuss equations for different types of squircles. We then build upon these 2D shapes to come-up with various 3D surfaces based on squircles.

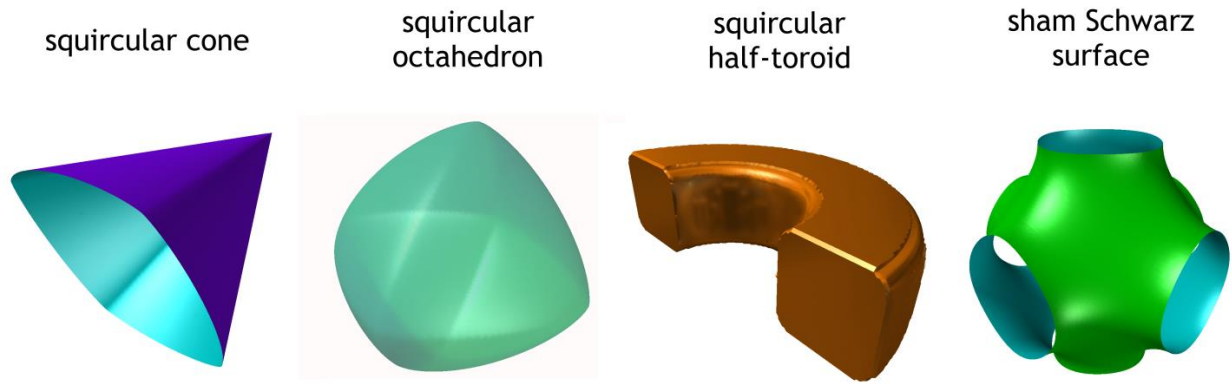

Figure 1: Some implicit surfaces with squircular cross sections.

## 1. Introduction

The square and the circle are among the most common shapes used by mankind. There are many 3D shapes that are derived from either the square or the circle as a base shape. Meanwhile, the squircle [Weisstein 2012] [Fong 2019] is a hybrid shape between the square and the circle. It is natural to ask whether the squircle can be used to replace the square or the circle to produce novel 3D shapes. This paper will examine the possibilities. Figure 1 shows some examples of our results. All of the shapes in Figure 1 have squircular cross sections. For example, the squircular cone has its base shape as a squircle instead of a circle. Also, the squircle is substituted for the square within the squircular octahedron and the squircular toroid. Lastly, the orifices of the sham Schwarz surface are all squircles.

Although we will approach the squircle from a purely mathematical viewpoint in this paper, we would like to mention that the squircle has been used in numerous applications. These diverse applications include microwave communications [Linz 2019], wind turbines [Roccia 2021], heat transfer [Lienhard 2019], rocketry [Wang 2021], and human psychology [Jaso 2021].

## 2. Different Types of Squircles

There are many different types of squircles, with each uniquely defined by its mathematical equation. In fact, any shape that is able to interpolate between the square and the circle can be considered as a squircle. In this paper, we will discuss 7 different types of squircles. We will provide a corresponding equation for each of these squircles centered at the origin. We will also discuss two parameters that come with each squircle. The $1^{st}$ is an interpolating variable which determines whether the shape is a circle or a square. The $2^{nd}$ is a scaling variable which determines the size of the squircle.

The most famous squircle is the superellipse, also known as the Lamé curve. The equation for the superellipse with no eccentricity is

$$|x|^P + |y|^P = r^P$$

There are two parameters in this equation: *p* and *r*. The power parameter *p* is an interpolating variable that allows one to blend the circle with the square. The scaling parameter *r* specifies the size of the shape. Incidentally, the superellipse actually contains two different families of squircles from its single equation. We shall refer to these two shapes as the *Lamé upper squircle* and the *Lamé lower squircle*.

## 2.1 Lamé Upper Squircle

The *Lamé upper squircle* is the resulting shape for $p \in [2,+\infty)$. When $p = 2$, the equation produces a circle with radius *r*. As $p \to +\infty$, the equation produces a square with a side length of *2r*. In between, the equation produces a smooth planar curve that blends the circle with the square. These are shown in Figure 2.

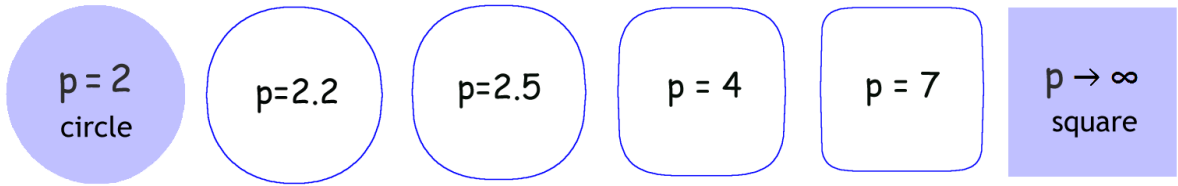

Figure 2: The Lamé upper squircle $|x|^P + |y|^P = r^P$ for $p \in [2,+\infty)$

The Lamé upper squircle can only approximate the square. It requires an infinite exponent in order to fully realize the square. Moreover, since the equation for the Lamé curve has unbounded exponents, it is unwieldy and difficult to manipulate algebraically. One of the main reasons for this paper is to find alternative shapes that are similar to the Lamé upper squircle but do not involve infinity.

The Lamé upper squircle has a 3D counterpart, which is an intermediate shape between the sphere and the cube. This has the equation:

$$|x|^P + |y|^P + |z|^P = r^P$$

The parameters in this equation are directly analogous to those in the 2D case. When *p=2*, the equation produces a sphere with radius *r*. As $p \to +\infty$, the equation produces a cube with a side length of *2r*. In between, the equation produces a smooth surface that blends the sphere with the cube. These are shown in Figure 3.

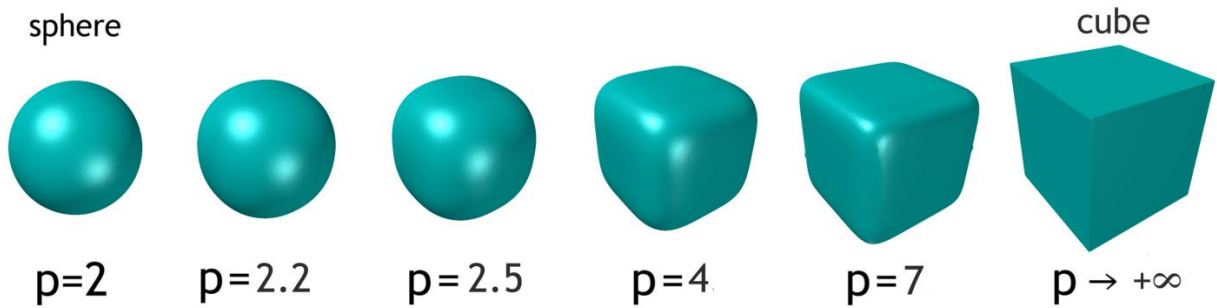

Figure 3: 3D counterpart of the Lamé upper squircle

For most of the 3D figures in this paper, we used the Marching Cubes algorithm [Lorensen 1987] [Bourke 1994] for converting implicit surfaces into triangular meshes which can then be displayed on a computer. The Marching Cubes algorithm is a popular technique in computer graphics for visualizing implicit surfaces. It has an adjustable parameter that allows for more accurate representation of the implicit surface at the cost of increasing triangle count.

## 2.2 Lamé Lower Squircle

The *Lamé lower squircle* is the resulting shape for $p \in [1,2]$. When $p = 1$, the equation produces a tilted square with side length of $r\sqrt{2}$. When $p = 2$, the equation produces a circle with radius $r$. In between, the equation produces a smooth planar curve that blends the circle with the square. These are shown in Figure 4. Notice that the square is tilted by 45°

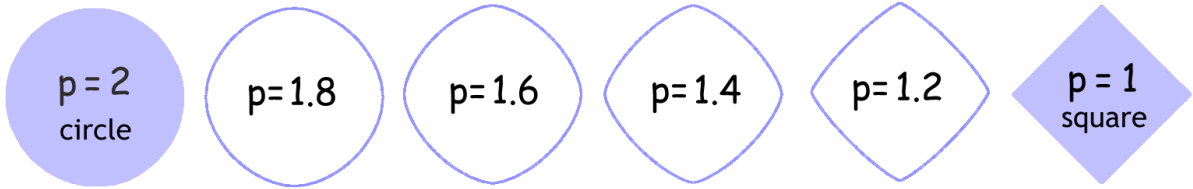

Figure 4: The Lamé lower squircle $|x|^P + |y|^P = r^P$ for $p \in [1,2]$

The Lamé lower squircle is algebraically simpler than the Lamé upper squircle because its exponent does not involve infinity. However, it produces a square that is tilted by 45°, which makes it less appealing for some applications. Also, like the Lamé upper squircle, its equation involves exponents that are not necessarily integers. A polynomial equation would be much simpler. In the next sub-section, we will discuss a squircle with that property.

The Lamé lower squircle has a 3D counterpart, which is an intermediate shape between the sphere and the octahedron. The equation for this is

$$|x|^P + |y|^P + |z|^P = r^P$$

with $p \in [1,2]$. When $p = 1$, the equation produces a regular octahedron with side length of $r\sqrt{2}$. When $p = 2$, the equation produces a sphere with radius $r$. In between, the equation produces a smooth surface that blends the octahedron with the sphere. These are shown in Figure 5.

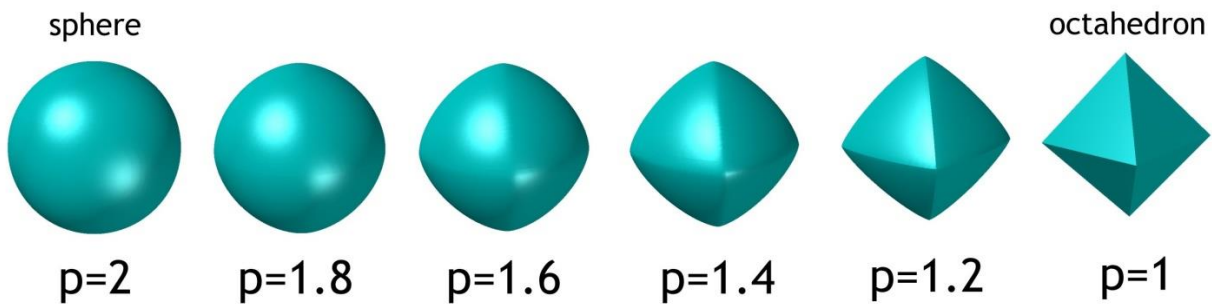

Figure 5: 3D counterpart of the Lamé lower squircle

Note that the regular octahedron is the dual of the cube, so it is a valid counterpart of the square in three dimensions. Also, it is easy to check that the octahedron has the implicit equation $|x| + |y| + |z| = r$.

## 2.3 Fernandez-Guasti Squircle

In 1992, Manuel Fernandez-Guasti discovered another type of squircle [Fernandez-Guasti 1992] [Fernandez-Guasti 2005]. This shape is a quartic plane curve, which makes it algebraically much simpler than the Lamé squircles. The equation for this shape is

$$x^2 + y^2 - \frac{s^2}{r^2} x^2 y^2 = r^2$$

There are two parameters for this equation: $s$ and $r$. The squareness parameter $s$ is the interpolating variable. When $s = 0$, the equation produces a circle with radius $r$. When $s = 1$, the equation produces a square with a side length of $2r$. In between, the equation produces a smooth planar curve that resembles both the circle and the square.

We have thoroughly covered and discussed the Fernandez-Guasti squircle in a previous paper [Fong 2018]. Hence, we only mention it briefly for this paper. The 3D counterpart [Fong 2018] for the Fernandez-Guasti squircle is called the *sphube* and has the equation:

$$x^2 + y^2 + z^2 - \frac{s^2}{r^2} x^2 y^2 - \frac{s^2}{r^2} y^2 z^2 - \frac{s^2}{r^2} x^2 z^2 + \frac{s^4}{r^4} x^2 y^2 z^2 = r^2$$

## 2.4 Frantz Squircle

In 2018, Marc Frantz discovered another type of squircle [Frantz 2018]. His squircle uses parametric equations instead of an implicit equation relating $x$ and $y$ on the Cartesian plane. The parametric equations are

$$x = \frac{r \tanh(s \cos t)}{\tanh s}$$

$$y = \frac{r \tanh(s \sin t)}{\tanh s}$$

The parameter $t$ is the standard variable of the parametric equation. It can take on any value between 0 and $2\pi$ to span the full squircular curve. There are two other variables for this equation: $s$ and $r$. The squareness variable $s$ allows the shape to interpolate between the circle and the square. As $s \to 0$, the equation produces a circle with radius $r$. When $s \to +\infty$, the equation produces a square with a side length of $2r$. In between, the parametric equation produces a smooth planar curve that is a hybrid of the circle and the square. This is shown in Figure 6.

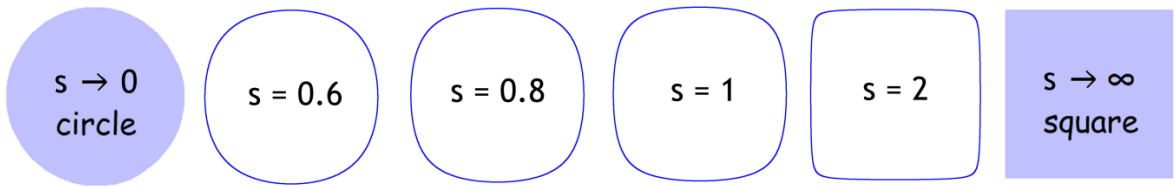

Figure 6: The Frantz squircle

Note that the Frantz squircle appears very similar to the Lamé upper squircle. This is true for most other squircles discussed in this paper. Even though the equations for these curves are distinct and dissimilar, the plotted squircular shapes share an uncanny resemblance. As a matter of fact, it can be difficult to tell these shapes apart from each other at varying squareness values.

## 2.5 Complex Squircle

There is yet another type of squircle defined in terms of complex variables [Fong 2019][Fong 2021]. The equations are quite complicated and involve elliptic integrals. We have decided not to include the equations in this paper, but provide a link in the references.

# 3. The Periodic Squircle

In this section, we will introduce a new type of squircle. This squircle is doubly-periodic in the Cartesian plane, hence the name. The equation for the *periodic squircle* is

$$\cos\left(\frac{s\pi x}{2r}\right) \cos\left(\frac{s\pi y}{2r}\right) = \cos\left(\frac{s\pi}{2}\right)$$

There are two parameters for this equation: *s* and *r*. The squareness parameter *s* is the interpolating variable. As $s \to 0$, the equation produces a circle with radius *r*. When $s = 1$, the equation produces a square with a side length of *2r*. In between, the equation produces a smooth planar curve that resembles both the circle and the square. This is shown in Figure 7. Note that there is a degeneracy in the equation when $s = 0$, because it reduces to $0 = 0$, which cannot be plotted.

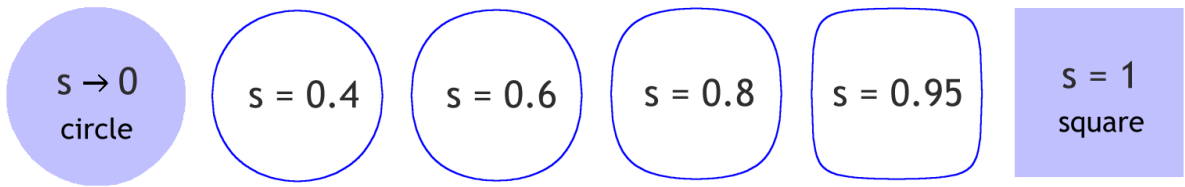

Figure 7: The periodic squircle

In Figure 7, we primarily focused in the region near the origin where $-r \leq x \leq r$ and $-r \leq y \leq r$. The implicit equation for the periodic squircle can actually have points (x,y) outside this region of interest. In fact, since this shape is doubly-periodic in the Cartesian plane, it repeats indefinitely along both the x and y directions. To illustrate this, the diagram in Figure 8 shows the periodic squircle plotted to extend outside the main region of interest. Observe that the squircular shape repeats in periodic intervals across the Cartesian plane.

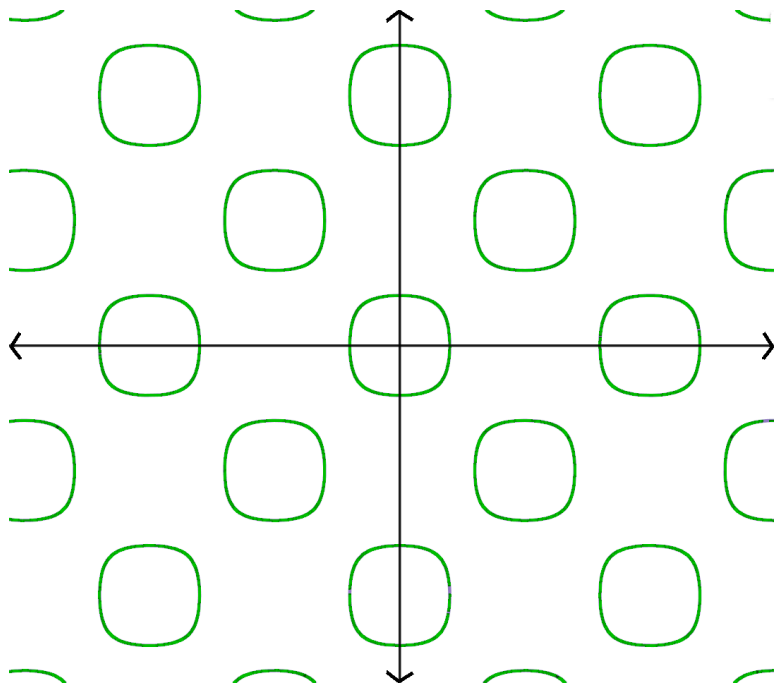

Figure 8: The periodic squircle repeats indefinitely on the Cartesian plane

When *s* = *1*, the periodic squircles join together form an infinite square grid. At the center of this grid is a single square within the region $-r \leq x \leq r$ and $-r \leq y \leq r$. This is shown in Figure 9. Observe that the square repeats with a period of *2r* in both the *x* and *y* directions.

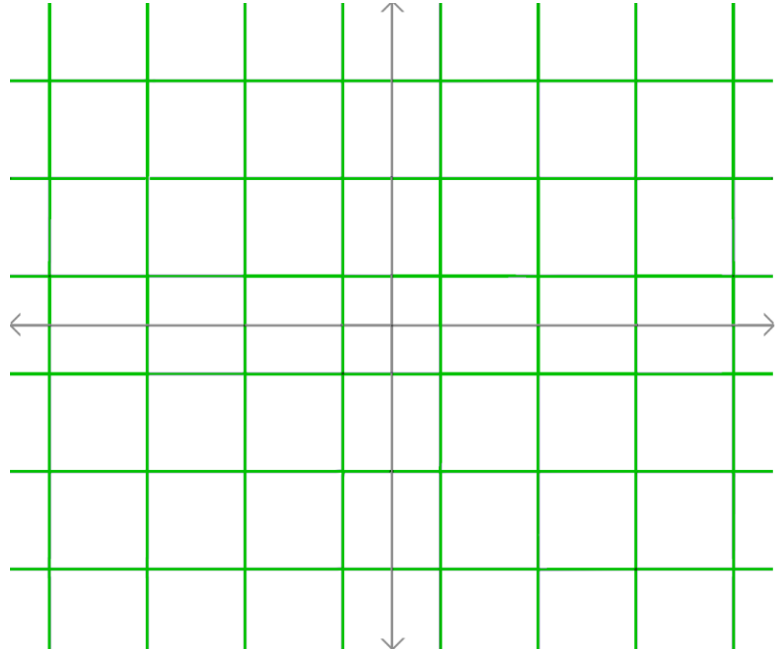

Figure 9: The periodic squircle becomes an infinite square grid when *s* = *1*

The periodic squircle has a 3D counterpart, which is an intermediate shape between the sphere and the cube. This is shown in Figure 10. This counterpart has the implicit equation:

$$\cos(\frac{s\pi x}{2r}) \; \cos(\frac{s\pi y}{2r}) \; \cos(\frac{s\pi z}{2r}) \; = \; \cos(\frac{s\pi}{2})$$

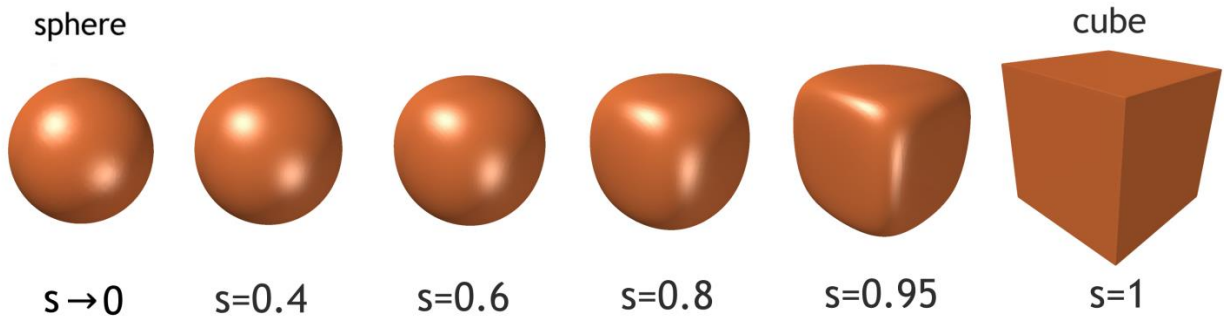

Figure 10: 3D counterpart of the periodic squircle

The parameters in this equation are directly analogous to those in the 2D case. As $s \to 0$, the equation produces a sphere with radius *r*. When *s* = *1*, the equation produces a cube with a side length of *2r*. In between, the equation produces a smooth closed surface that blends the sphere with the cube. These are shown in Figure 10. Note that there is a degeneracy in the equation when *s* = *0*, because it reduces to *0* = *0*, which cannot be rendered in 3D Cartesian space.

In Figure 10, we primarily focused in the region near the origin where $-r \leq x \leq r$ and $-r \leq y \leq r$ and $-r \leq z \leq r$. In analogy to the periodic squircle in 2D, the implicit equation for its 3D counterpart can have points outside the main region of interest. In fact, it is triply-periodic in Cartesian space. This is shown in Figure 11.

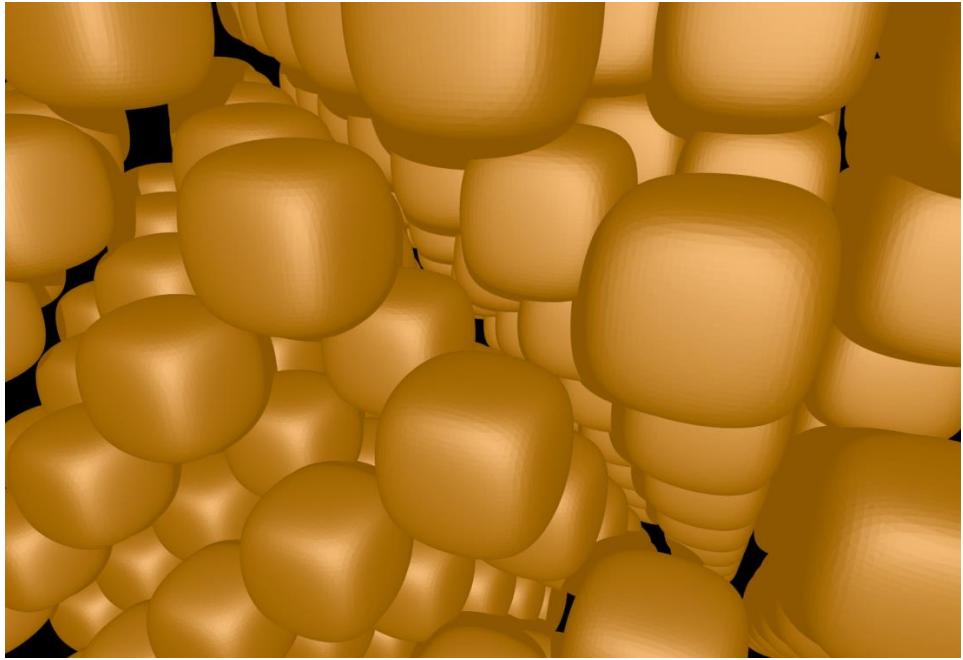

Figure 11: The 3D counterpart is triply-periodic in Cartesian space

## 4. A Powerful Generalization of the Periodic Squircle

There is a generalization of the periodic squircle in which there is an extra parameter that controls the radius of the circle when $s \to 0$. The equation for this generalized periodic squircle is

$$\cos\left(\frac{s\pi x}{2r}\right) \cos\left(\frac{s\pi y}{2r}\right) = \left[\cos\left(\frac{s\pi}{2}\right)\right]^p$$

This squircle behaves just like the periodic squircle except that when $s \to 0$, the equation produces a circle with radius $r\sqrt{p}$. When $s = 1$, the equation still produces a square with a side length of $2r$. The independent $p$ exponent allows for the circle and the square to have very different sizes, yet still have a full spectrum of intermediate squircles between the two. This is illustrated in Figure 12 which plots the generalized periodic squircle with $p=9$.

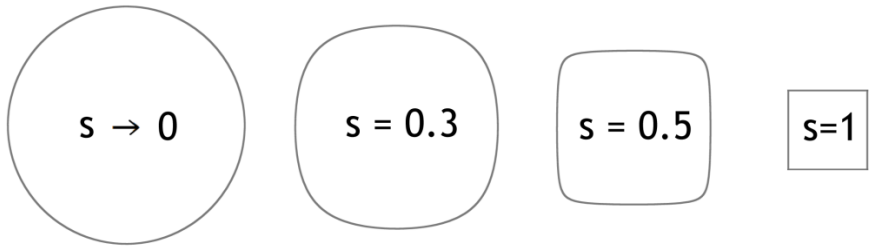

Figure 12: The generalized periodic squircle has an extra power parameter that controls the size of the circle

Observe in Figure 12 that generalized periodic squircle not only interpolates the shape between the circle and the square but also interpolates between the sizes of the two shapes. By introducing the parameter $p$, we can control the size of the circle independently from the size of the square.

The periodic squircle can be observed interactively online using the Desmos graphing calculator. Please follow the hyperlink below to run a demo with interactive sliders for control the different parameters of this squircle.
*https://www.desmos.com/calculator/cnmjtd2dwr*

Of course, the generalized periodic squircle easily extends to 3D with the following equation.

$$\cos\left(\frac{s\pi x}{2r}\right) \cos\left(\frac{s\pi y}{2r}\right) \cos\left(\frac{s\pi z}{2r}\right) = \left[\cos\left(\frac{s\pi}{2}\right)\right]^p$$

There is a another variant of the periodic squircle which we will only briefly mention here

$$\cos\left(\frac{s\pi x}{2r}\right) \cos\left(\frac{s\pi y}{2r}\right) = \left[1 - \sin\left(\frac{s^2\pi}{2}\right)\right]^p$$

# 5. The Infinite Square Grid

In this section, we will derive an implicit equation for an infinite square grid on the Cartesian plane. This equation is not necessarily unique, but it will be useful in the next section where we will show that the periodic squircle is indeed a squircle.

## 5.1 Horizontal Lines

Consider the equation $cos\ y = 0$. This equation implies that $y = (2n + 1)\frac{\pi}{2}$ for all integers $n$. Similarly, the equation $cos\left(\frac{\pi y}{2}\right) = 0$ implies that $y = 2n + 1$ for all integers $n$. If the equation is graphed in the Cartesian plane, it will consist of an infinitude of equally-spaced horizontal lines. This is shown in Figure 12.

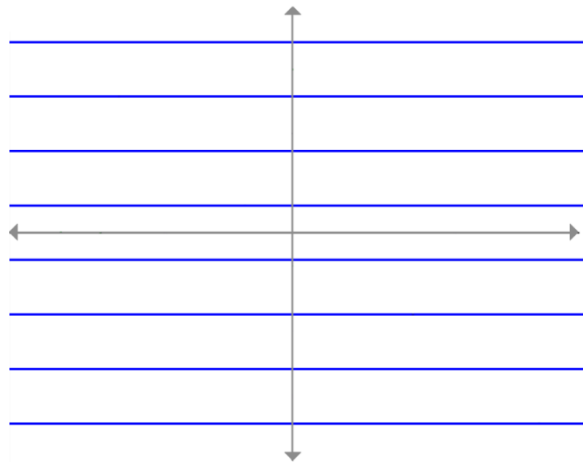

Figure 12: The graph of $cos\left(\frac{\pi y}{2}\right) = 0$

## 5.2 Vertical Lines

Consider the equation $cos\ x = 0$. This equation implies that $x = (2n + 1)\frac{\pi}{2}$ for all integers $n$. Similarly, the equation $cos\left(\frac{\pi x}{2}\right) = 0$ implies that $x = 2n + 1$ for all integers $n$. If the equation is graphed in the Cartesian plane, it will consist of an infinitude of equally-spaced vertical lines. This is shown in Figure 13.

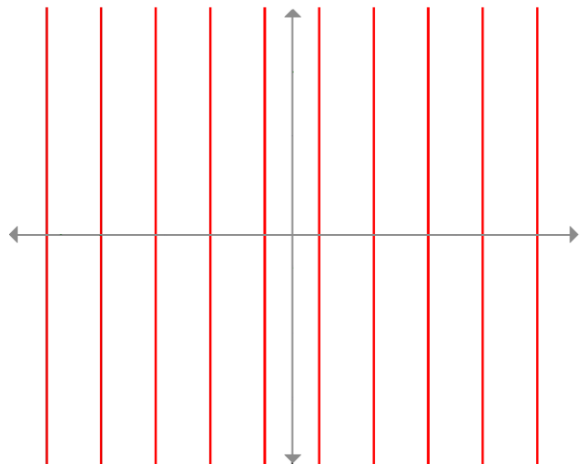

Figure 13: The graph of $cos\left(\frac{\pi x}{2}\right) = 0$

## 5.3 Overlaid Square Grid

It is possible to superimpose the horizontal lines with the vertical lines to get an infinite square grid on the Cartesian plane. This can be done by simply multiplying our equations for horizontal and vertical lines to get the implicit equation

$$\cos\left(\frac{\pi x}{2}\right)\cos\left(\frac{\pi y}{2}\right) = 0$$

This equation is graphed in Figure 14 below.

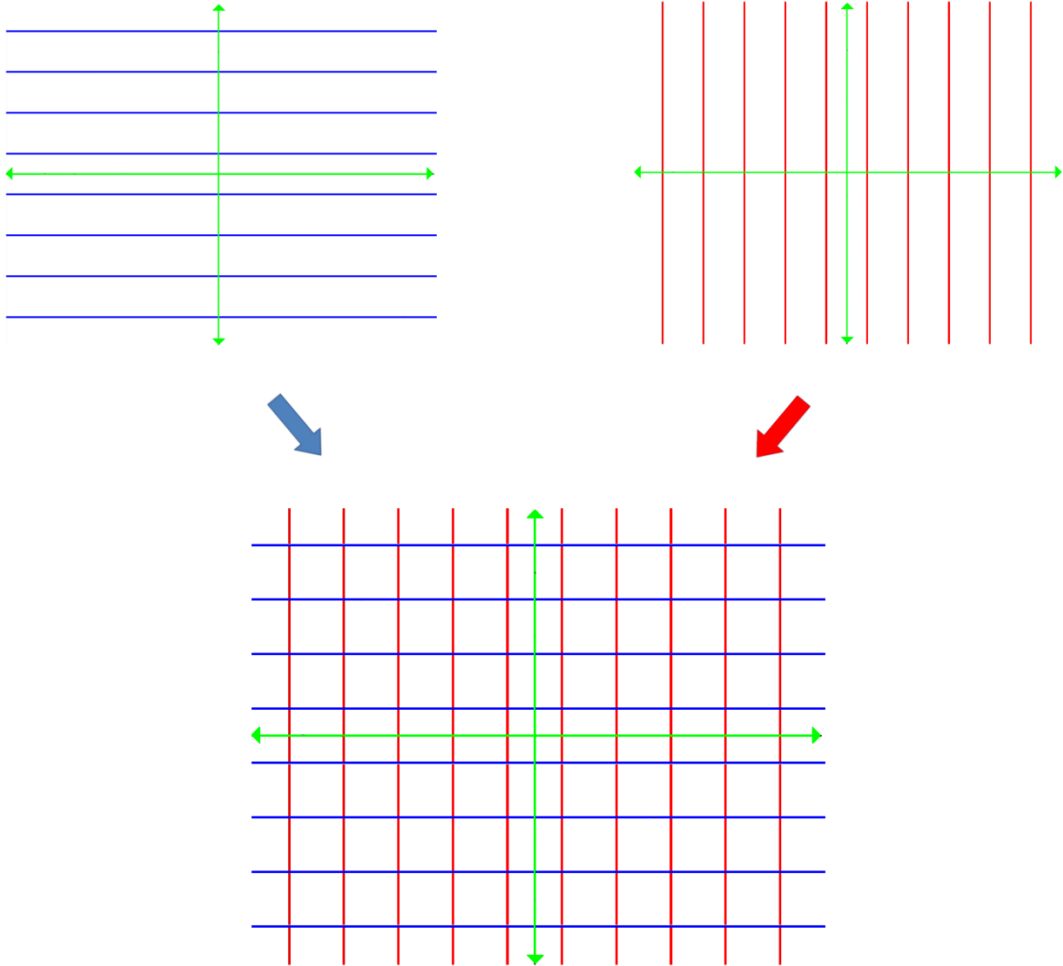

Figure 14: The graph of $\cos\left(\frac{\pi x}{2}\right)\cos\left(\frac{\pi y}{2}\right) = 0$ comes from superimposing horizontal & vertical lines

Thus, we have derived an implicit equation for the infinite square grid on the Cartesian plane. If we want to incorporate a scale factor *r* that gives the square grid a side length of *2r*, the implicit equation can be modified to

$$\cos\left(\frac{\pi x}{2r}\right)\cos\left(\frac{\pi y}{2r}\right) = 0$$

Using a similar approach, one can easily extend this result to 3D and get an implicit equation for an infinite cubic grid in Cartesian space, which is

$$\cos\left(\frac{\pi x}{2r}\right)\cos\left(\frac{\pi y}{2r}\right)\cos\left(\frac{\pi z}{2r}\right) = 0$$

An interesting phase-shifted variant of the square grid arises when the intersection points between horizontal and vertical lines are all lattice points with integer coordinates *(x,y)*, where $x \in \mathbb{Z}$ and $y \in \mathbb{Z}$. The implicit equation for this variant is: $\sin(\pi x)\sin(\pi x) = 0$

## 6. Proof Outline for the Periodic Squircle

In this section, we will show that the generalized periodic squircle is indeed a squircle, i.e. an intermediate shape between the square and the circle. Without loss of generality, let $\omega = s\pi$. The equation for the generalized periodic squircle then reduces to

$$\cos(\tfrac{\omega x}{2r}) \cos(\tfrac{\omega y}{2r}) = \left[\cos(\tfrac{\omega}{2})\right]^p$$

The equation for this squircle can be manipulated to get isolated expressions for both $x$ and $y$.

$$x = \frac{2r}{\omega} \cos^{-1} \frac{\left[\cos(\tfrac{\omega}{2})\right]^p}{\cos \tfrac{\omega y}{2r}} \qquad\qquad y = \frac{2r}{\omega} \cos^{-1} \frac{\left[\cos(\tfrac{\omega}{2})\right]^p}{\cos \tfrac{\omega x}{2r}}$$

### 6.1 Circular Case when $s \to 0$

When $\to 0$, we also get $\omega \to 0$. In order to demonstrate that this curve is a type of squircle, we have to first show that as $\omega$ goes to zero, the equation becomes circular with a radius of $r\sqrt{p}$. Note that if we just set $\omega = 0$, the implicit equation becomes degenerate of the form $0 = 0$, which cannot be plotted. In order to avoid this degeneracy, we need to take the limit as $\omega$ goes to zero instead. In other words, we need to show that

$$\lim_{\omega \to 0} y = \sqrt{r^2 p - x^2}$$

which is equivalent to the equation for the circle with radius $r\sqrt{p}$ centered at the origin, i.e.

$$x^2 + y^2 = r^2 p$$

Substituting the expression for x, we get this equation to be proved

$$\lim_{\omega \to 0} \frac{2r}{\omega} \cos^{-1} \frac{\left[\cos(\tfrac{\omega}{2})\right]^p}{\cos \tfrac{\omega x}{2}} = \sqrt{r^2 p - x^2}$$

Using L'Hôpital's rule, we can simplify the limit by differentiating both the numerator and denominator with respect to $\omega$ in order to get

$$\lim_{\omega \to 0} \frac{2r}{\omega} \cos^{-1} \frac{\left[\cos(\tfrac{\omega}{2})\right]^p}{\cos \tfrac{\omega x}{2}} = \lim_{\omega \to 0} \frac{rp\left[\cos \tfrac{\omega}{2}\right]^{p-1} \sin \tfrac{\omega}{2} \sec \tfrac{\omega x}{2} - x\left[\cos \tfrac{\omega}{2}\right]^p \tan \tfrac{\omega x}{2} \sec \tfrac{\omega x}{2}}{\sqrt{1 - \left[\cos \tfrac{\omega}{2}\right]^{2p} \left[\sec \tfrac{\omega x}{2}\right]^2}}$$

Furthermore, we can do a Taylor series expansion of the numerator and denominator at $\omega = 0$ to further simplify.

For the numerator:

$$rp\left[\cos \tfrac{\omega}{2}\right]^{p-1} \sin \tfrac{\omega}{2} \sec \tfrac{\omega x}{2} - x\left[\cos \tfrac{\omega}{2}\right]^p \tan \tfrac{\omega x}{2} \sec \tfrac{\omega x}{2}$$
$$= \frac{\omega}{2r}(r^2 p - x^2) - \frac{\omega^3}{48r^3}(pr^4 + 3p^2 r^4 - 3p^2 r^2 - 6px^2 r^2 + 5x^4) + O(\omega^5)$$

where $O(\omega^5)$ represents an infinite sum of higher order polynomial terms in $\omega$ with exponents $\geq 5$.

For the term inside the square root in the denominator:

$$1 - \left[\cos\frac{\omega}{2}\right]^{2p} \left[\sec\frac{\omega x}{2}\right]^2 = \frac{\omega^2}{4r^2}(r^2p - x^2) - \frac{\omega^4}{96r^4}(5pr^4 + 3p^2r^4 - 6x^2r^2 - 6px^2r^2 + 4x^4) + O(\omega^6)$$

where $O(\omega^6)$ represents an infinite sum of higher order polynomial terms in $\omega$ with exponents $\geq 6$.

Substituting in the Taylor series expansions for the numerator and denominator, we get

$$\lim_{\omega \to 0} \frac{2r}{\omega} \cos^{-1} \frac{\left[\cos(\frac{\omega}{2})\right]^p}{\cos\frac{\omega x}{2}}$$

$$= \lim_{\omega \to 0} \frac{\frac{\omega}{2r}(r^2p - x^2) - \frac{\omega^3}{48r^3}(pr^4 + 3p^2r^4 - 3p^2r^2 - 6px^2r^2 + 5x^4) + O(\omega^5)}{\sqrt{\frac{\omega^2}{4r^2}(r^2p - x^2) - \frac{\omega^4}{96r^4}(5pr^4 + 3p^2r^4 - 6x^2r^2 - 6px^2r^2 + 4x^4) + O(\omega^6)}}$$

$$= \lim_{\omega \to 0} \frac{\frac{\omega}{2r}(r^2p - x^2) - \frac{\omega^3}{48r^3}(pr^4 + 3p^2r^4 - 3p^2r^2 - 6px^2r^2 + 5x^4) + O(\omega^5)}{\frac{\omega}{2r}\sqrt{r^2p - x^2}\sqrt{1 - \frac{\omega^2}{24r^2}\frac{(5pr^4 + 3p^2r^4 - 6x^2r^2 - 6px^2r^2 + 4x^4)}{r^2p - x^2} + O(\omega^4)}}$$

$$= \lim_{\omega \to 0} \frac{\frac{\omega}{2r}(r^2p - x^2)\left(1 + \frac{\omega^2}{24r^2}\frac{(pr^4 + 3p^2r^4 - 3p^2r^2 - 6px^2r^2 + 5x^4)}{(r^2p - x^2)} + O(\omega^4)\right)}{\frac{\omega}{2r}\sqrt{r^2p - x^2}\sqrt{1 - \frac{\omega^2}{24r^2}\frac{(5pr^4 + 3p^2r^4 - 6x^2r^2 - 6px^2r^2 + 4x^4)}{r^2p - x^2} + O(\omega^4)}}$$

$$= \sqrt{r^2p - x^2} \lim_{\omega \to 0} \frac{\left(1 + \frac{\omega^2}{24r^2}\frac{(pr^4 + 3p^2r^4 - 3p^2r^2 - 6px^2r^2 + 5x^4)}{(r^2p - x^2)} + O(\omega^4)\right)}{\sqrt{1 - \frac{\omega^2}{24r^2}\frac{(5pr^4 + 3p^2r^4 - 6x^2r^2 - 6px^2r^2 + 4x^4)}{r^2p - x^2} + O(\omega^4)}}$$

$$= \sqrt{r^2p - x^2}$$

**6.2 Square Case when** *s =1*

This case is very simple because of our derivations from the previous section. Just substitute *s=1* into the equation of the periodic squircle and simplify to get the implicit equation

$$\cos(\frac{\pi x}{2r}) \cos(\frac{\pi y}{2r}) = 0$$

We already know from Section 5 that this implicit equation produces an infinite square grid in the Cartesian plane. At the center of the grid is a square with side length of *2r*. This shows that the periodic squircle is a square near the origin when *s = 1*.

# 7. The Oblique Squircle

In this section, we introduce another type of squircle that is also doubly-periodic in the Cartesian plane. The implicit equation for the *oblique squircle* is

$$\cos\left(\frac{s\pi x}{r}\right) + \cos\left(\frac{s\pi y}{r}\right) = 1 + \cos(s\pi)$$

There are two parameters for this equation: *s* and *r*. The squareness parameter *s* is the interpolating variable. As *s* → 0, the equation produces a circle with radius *r*. When *s* =1, the equation produces a square with a side length of *r*√2. In between, the equation produces a smooth planar curve that a hybrid of the circle and the square. We show the oblique squircle at varying squareness in Figure 15. Notice that the square is tilted by 45°. There is a degeneracy in the equation when *s* = 0, because it reduces to 2 = 2, which cannot be plotted. The oblique squircle has a strong resemblance to the Lamé lower squircle.

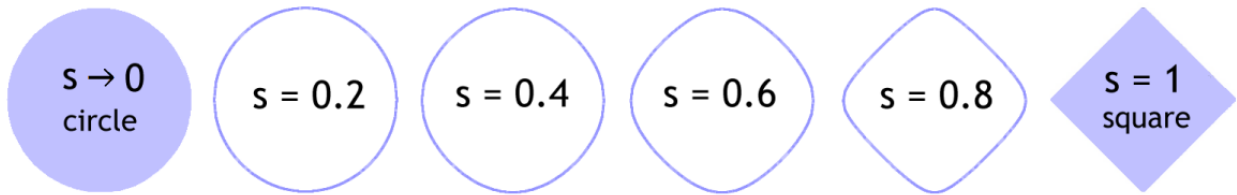

Figure 15: The oblique squircle

In Figure 15, we primarily focused in the region near the origin where -*r* ≤ *x* ≤ *r* and -*r* ≤ *y* ≤ *r*. The implicit equation for the oblique squircle can actually have points outside this region of interest. In fact, since this shape is doubly-periodic in the Cartesian plane, it repeats indefinitely along both the x and y directions. To illustrate this, the diagram shown in Figure 16 shows the oblique squircle plotted to extend outside the main region of interest. Observe that the squircular shape repeats in periodic intervals across the Cartesian plane.

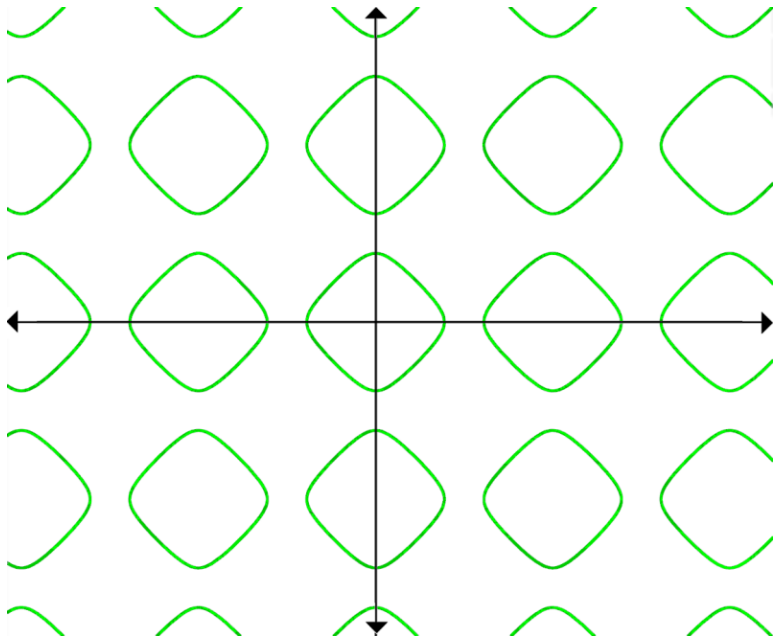

Figure 16: The oblique squircle repeats indefinitely on the Cartesian plane

When *s = 1*, the oblique squircles join together form a slanted square grid. At the center of this grid is a single square tilted by 45°. This is shown in Figure 17. Observe that this tilted square repeats with a period of *2r* in both the *x* and *y* directions.

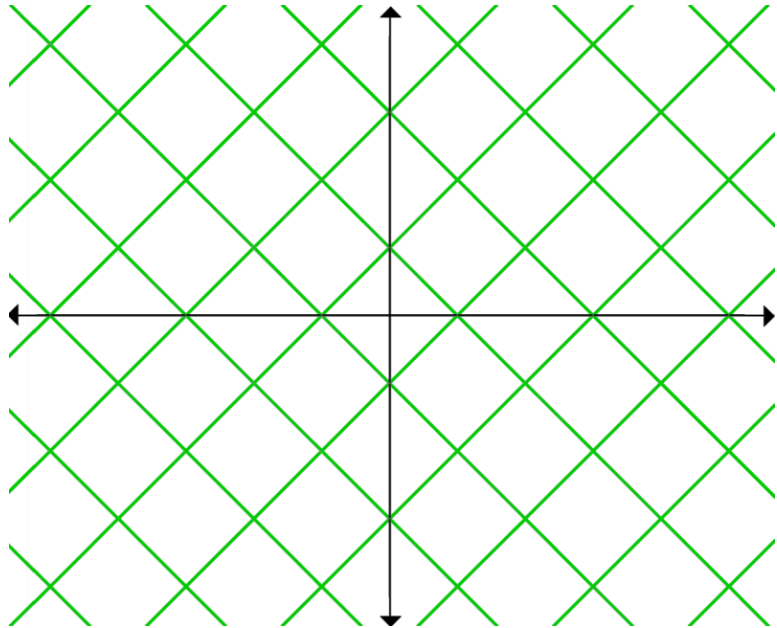

Figure 17: The oblique squircle becomes a slanted square grid when *s = 1*

Just like the Lamé lower squircle, the oblique squircle can be extended to 3D. However, the corresponding 3D shape to the square will not be an octahedron or a cube. Instead, it is a shape that resembles the regular octahedron which we shall name as the *sham octahedron*. The equation for the 3D counterpart is

$$\cos\left(\frac{s\pi x}{r}\right) + \cos\left(\frac{s\pi y}{r}\right) + \cos\left(\frac{s\pi z}{r}\right) = 2 + \cos(s\pi)$$

This implicit equation becomes a sphere with radius *r* as *s* → 0. When *s = 1*, the implicit equation produces the sham octahedron. In between, the implicit equation produces an octahedral shape that interpolates between the sphere and the sham octahedron. The other parameter *r* provides a scale factor. The octahedral shape is shown in Figure 18.

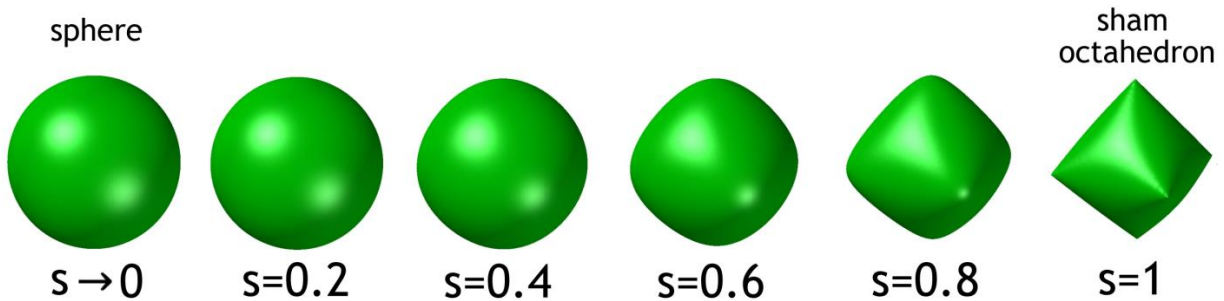

Figure 18: 3D counterpart of the oblique squircle

Although the sham octahedron strongly resembles the regular octahedron, it is not even a polyhedron. The sham octahedron does not have flat polygonal faces.

In Figure 18, we primarily focused in the region near the origin where $-r \leq x \leq r$ and $-r \leq y \leq r$ and $-r \leq z \leq r$. In analogy to the oblique squircle in 2D, the implicit equation for its 3D counterpart can have points outside the main region of interest. In fact, the counterpart is triply-periodic in Cartesian space. This is shown in Figure 19.

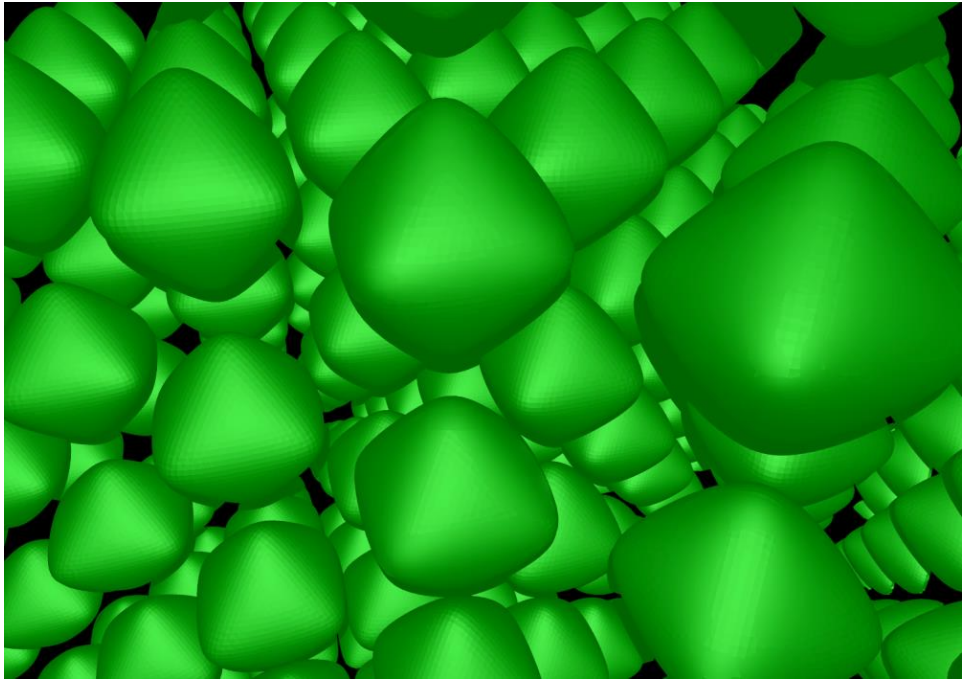

Figure 19: The 3D counterpart is triply periodic in Cartesian space

# 8. Cartesian Rotation by 45°

In this section, we will review rotation by 45° in the Cartesian coordinate system. We need to do this in order to show the relation between the oblique squircle and the periodic squircle in the next section.

Recall from linear algebra the formula for rotation in the Cartesian coordinate system. Let *(x,y)* be a point and *(u,v)* be its rotated counterpart at an angle of $\theta$, counterclockwise from the positive x-axis. The rotation formula relating *(u,v)* with *(x,y)* is

$$\begin{bmatrix} u \\ v \end{bmatrix} = \begin{bmatrix} \cos\theta & -\sin\theta \\ \sin\theta & \cos\theta \end{bmatrix} \begin{bmatrix} x \\ y \end{bmatrix}$$

For a clockwise rotation angle of 45°, we have $\theta = -45$. The equation then simplifies to

$$\begin{bmatrix} u \\ v \end{bmatrix} = \frac{1}{\sqrt{2}} \begin{bmatrix} 1 & 1 \\ -1 & 1 \end{bmatrix} \begin{bmatrix} x \\ y \end{bmatrix}$$

with an inverse relation as

$$\begin{bmatrix} x \\ y \end{bmatrix} = \frac{1}{\sqrt{2}} \begin{bmatrix} 1 & -1 \\ 1 & 1 \end{bmatrix} \begin{bmatrix} u \\ v \end{bmatrix}$$

Thus, we have these two main equations for clockwise rotation by 45°

$$x = \tfrac{1}{\sqrt{2}}(u - v)$$
$$y = \tfrac{1}{\sqrt{2}}(u + v)$$

In order to rotate a shape by 45° in the Cartesian coordinate system, we can simplify substitute these two *(u,v)* transformation equations for *x* and *y* in the implicit equation. We will now show some examples

**Example#1:** a sloped line $y = x$

Consider the implicit equation: $y = x$, which is a simple line with a slope of 1. This line has an angle of 45° with respect to positive x-axis. If we want to rotate this clockwise by an angle of 45°, we expect to get the horizontal line *y=0*. In order to rotate a shape by 45° in the Cartesian coordinate system, we can use the *(u,v)* transformation equations provided above.

$$y = x \quad \Rightarrow \quad \tfrac{1}{\sqrt{2}}(u + v) = \tfrac{1}{\sqrt{2}}(u - v)$$

This simplifies to

$$v = -v \quad \Rightarrow \quad v = 0$$

which is equivalent to the horizontal line: *y=0* when renamed back to the standard Cartesian plane.

**Example#2:** a hyperbola $y = 1/x$

Consider the implicit equation: $y = 1/x$, which a hyperbola. If we want to rotate this clockwise by an angle of 45°, we expect to still get an equation for a hyperbola. This rotation can be done by using the *(u,v)* transformation equations provided above.

$$y = 1/x \quad \Rightarrow \quad \tfrac{1}{\sqrt{2}}(u + v) = \frac{1}{\tfrac{1}{\sqrt{2}}(u - v)}$$

This simplifies to

$$\tfrac{1}{2}(u^2 - v^2) = 1$$

which is indeed a hyperbola when renamed back to the standard Cartesian plane. Recall that the canonical equation for the standard hyperbola in Cartesian coordinate system is

$$\frac{x^2}{a^2} - \frac{y^2}{b^2} = 1$$

Thus, our rotated hyperbola has $a = b = \sqrt{2}$

**Example#3:** Fernandez-Guasti squircle $x^2 + y^2 - \frac{s^2}{r^2} x^2 y^2 = r^2$

Consider the implicit equation for the Fernandez-Guasti squircle: $x^2 + y^2 - \frac{s^2}{r^2} x^2 y^2 = r^2$. If we want to rotate this clockwise by an angle of 45°, we can use the *(u,v)* transformation equations.

$$x^2 + y^2 - \frac{s^2}{r^2} x^2 y^2 = r^2$$

$$\Rightarrow \left[\tfrac{1}{\sqrt{2}}(u-v)\right]^2 + \left[\tfrac{1}{\sqrt{2}}(u+v)\right]^2 - \frac{s^2}{r^2} \left[\tfrac{1}{\sqrt{2}}(u-v)\right]^2 \left[\tfrac{1}{\sqrt{2}}(u+v)\right]^2 = r^2$$

$$\Rightarrow \tfrac{1}{2}(u-v)^2 + \tfrac{1}{2}(u+v)^2 - \frac{s^2}{4r^2}(u^2-v^2)^2 = r^2$$

$$\Rightarrow (u-v)^2 + (u+v)^2 - \frac{s^2}{2r^2}(u^2-v^2)^2 = 2r^2$$

$$\Rightarrow u^2 - 2uv + v^2 + u^2 + 2uv + v^2 - \frac{s^2}{2r^2}(u^2-v^2)^2 = 2r^2$$

$$\Rightarrow 2u^2 + 2v^2 - \frac{s^2}{2r^2}(u^2-v^2)^2 = 2r^2$$

$$\Rightarrow u^2 + v^2 - \frac{s^2}{4r^2}(u^2-v^2)^2 = r^2$$

The variables can be renamed back to the standard Cartesian plane to get an equation for the 45° rotated version of the Fernandez-Guasti squircle. The implicit equation is plotted in Figure 20

$$x^2 + y^2 - \frac{s^2}{4r^2}(x^2-y^2)^2 = r^2$$

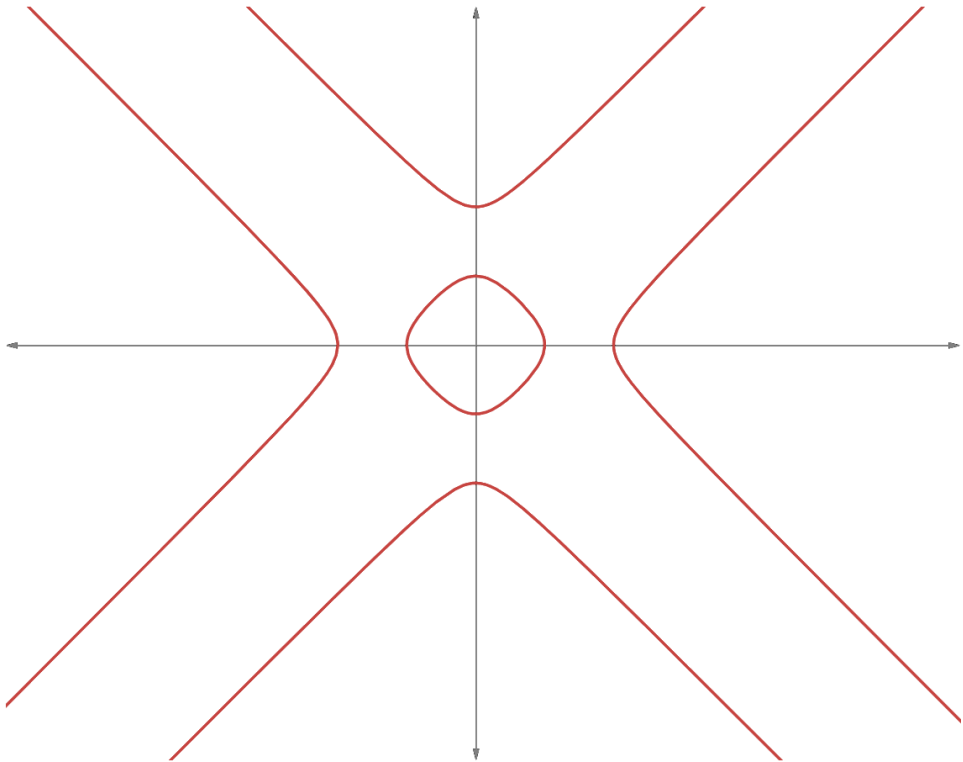

Figure 20: A 45° rotated version of the Fernandez-Guasti squircle at *s=0.8*

# 9. Connection Between the Periodic Squircle and the Oblique Squircle

In this section, we will show that the oblique squircle and the generalized periodic squircle are closely related. Specifically, we will show that oblique squircle is just an instance of the generalized periodic squircle after a rotation by 45 degrees. To show this, we start with the equation for the oblique squircle

$$\cos\left(\frac{s\pi x}{r}\right) + \cos\left(\frac{s\pi y}{r}\right) = 1 + \cos(s\pi)$$

Using the *(u,v)* transformation equations discussed in the previous section for a 45° clockwise rotation in the Cartesian plane

$$x = \frac{1}{\sqrt{2}}(u - v)$$
$$y = \frac{1}{\sqrt{2}}(u + v)$$

we get this rotated implicit equation:

$$\cos\left(\frac{s\pi(u - v)}{r\sqrt{2}}\right) + \cos\left(\frac{s\pi(u + v)}{r\sqrt{2}}\right) = 1 + \cos(s\pi)$$

Let $k = \frac{r}{\sqrt{2}}$ ⇒ $2k = r\sqrt{2}$ and substitute into the equation

$$\cos\left(\frac{s\pi(u - v)}{2k}\right) + \cos\left(\frac{s\pi(u + v)}{2k}\right) = 1 + \cos(s\pi)$$

Note that $r\sqrt{2}$ is just the side length of the square in the oblique squircle when *s=1*. The variable $k$ is just half of the side length of this square. This adjustment is needed for the squircle after a rotation of 45°.

Recall the product-to-sum formula from trigonometry, i.e.

$$2\cos A \cos B = \cos(A + B) + \cos(A - B)$$

or equivalently

$$2\cos(KA)\cos(KB) = \cos(K(A + B)) + \cos(K(A - B))$$

Therefore,

$$\cos\left(\frac{s\pi(u - v)}{2k}\right) + \cos\left(\frac{s\pi(u + v)}{2k}\right) = 1 + \cos(s\pi)$$

⇒ $$2\cos\left(\frac{s\pi u}{2k}\right)\cos\left(\frac{s\pi v}{2k}\right) = 1 + \cos(s\pi)$$

⇒ $$\cos\left(\frac{s\pi u}{2k}\right)\cos\left(\frac{s\pi v}{2k}\right) = \frac{1 + \cos(s\pi)}{2}$$

Now recall the half-angle formula from trigonometry, i.e.

$$\cos\frac{\theta}{2} = \sqrt{\frac{1 + \cos\theta}{2}} \qquad \Rightarrow \qquad \left(\cos\frac{\theta}{2}\right)^2 = \frac{1 + \cos\theta}{2}$$

Therefore,

$$\cos\left(\frac{s\pi u}{2k}\right)\cos\left(\frac{s\pi v}{2k}\right) = \frac{1 + \cos(s\pi)}{2}$$

⇒ $$\cos\left(\frac{s\pi u}{2k}\right)\cos\left(\frac{s\pi v}{2k}\right) = \left(\cos\frac{s\pi}{2}\right)^2$$

This last equation is just a specific equation at *p=2* for the generalized periodic squircle! Therefore, we can conclude that by just doing a rotation by 45° and a rescaling of the oblique squircle

$$\cos\left(\frac{s\pi x}{r}\right) + \cos(\frac{s\pi y}{r}) = 1 + \cos(s\pi)$$

we get an instance of the generalized periodic squircle

$$\cos\left(\frac{s\pi x}{2k}\right) \cos\left(\frac{s\pi y}{2k}\right) = \left(\cos\frac{s\pi}{2}\right)^2$$

In other words, the oblique squircle is an affine equivalent to the generalized periodic squircle at *p=2*. Using a similar derivation, we can generalize the oblique squircle just as we did for the periodic squircle. The generalized oblique squircle has the implicit equation

$$\cos\left(\frac{s\pi x}{r}\right) + \cos(\frac{s\pi y}{r}) = 2\left[\frac{1 + \cos(s\pi)}{2}\right]^p$$

with an extra parameter *p* that allows for control of the size of the circle when $s \to 0$ .

# 10. Overshoot in the Oblique Squircle

In this section, we will introduce the concept of *overshoot* within the oblique squircle. Much of this material is a retread of the paper "A Sham Schwarz Surface based on a Squircle" [Fong 2022], but with additional details. First, recall the implicit equation for the oblique squircle

$$\cos\left(\frac{s\pi x}{r}\right) + \cos\left(\frac{s\pi y}{r}\right) = 1 + \cos(s\pi)$$

Observe that the left hand side of the equation is bounded by this inequality

$$-2 \leq \cos\left(\frac{s\pi x}{r}\right) + \cos\left(\frac{s\pi y}{r}\right) \leq 2$$

whereas the right hand side of the equation is bounded by this inequality

$$0 \leq 1 + \cos(s\pi) \leq 2$$

So there is a discrepancy in the range of possible values between both sides of the equation for the oblique squircle. In order to account for the full range of values coming from the sum of two cosines, we introduce the concept of *overshoot* signified by the variable $h$. We then amend the implicit equation for the oblique squircle as

$$\cos\left(\frac{s\pi x}{r}\right) + \cos\left(\frac{s\pi y}{r}\right) = 1 + \cos(s\pi) - \lfloor s \rfloor h$$

One can consider the overshoot variable as an extra parameter independent from the squareness parameter. Intuitively, it is a parameter used to account for other possible shapes that arise from the sum of two cosines. The overshoot variable can have any value between zero and two, i.e. $h \in [0,2]$. Note that the amended implicit equation uses the floor function $\lfloor s \rfloor$ as a way to decouple the overshoot and squareness parameters. However, overshoot in the oblique squircle is only valid after the squareness parameter has reached a value of 1.

Figure 21 shows the effect of overshoot on the oblique squircle. The top row shows the oblique squircle with no overshoot. The bottom row shows it with increasing overshoot. At first glance, it appears that the curve becomes disconnected into 4 parts as overshoot increases. In reality, one has to take into account the doubly periodic nature of this shape in the Cartesian plane. Actually, overshoot introduces some kind of phase shift to the oblique squircle and then causes the squircle to recede into nothing at $h = 2$.

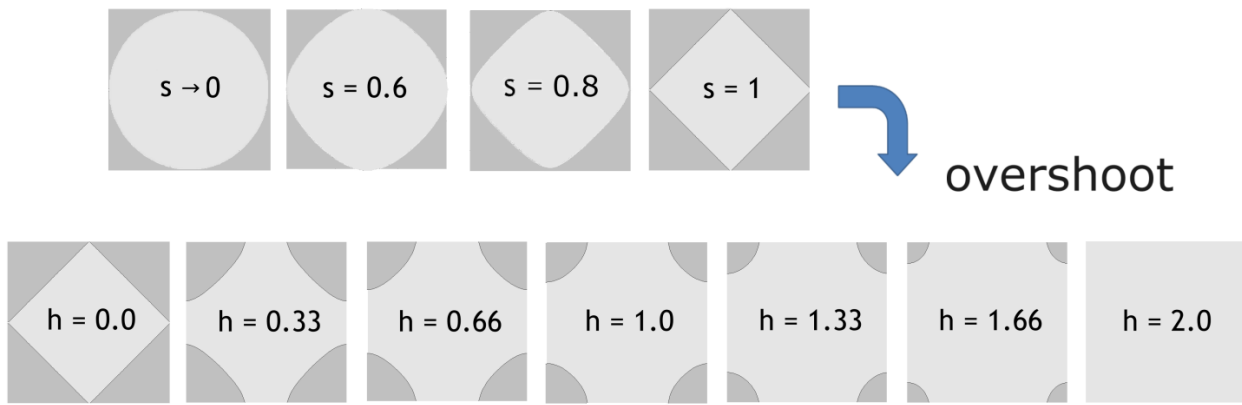

Figure 21: The oblique squircle can overshoot beyond the square

We strongly encourage the reader to observe the overshoot parameter interactively online via the Desmos graphing calculator. Desmos comes with interactive sliders that allow the user to observe how the squareness and overshoot parameters affect the appearance of the shape. Desmos also comes with a dynamic pan & zoom user interface that can showcase the twofold periodicity of the oblique squircle.

*https://www.desmos.com/calculator/23njdxyut9*

Figure 22 shows the graph of the oblique squircle with overshoot beyond our main region of interest. The extended 2D plot showcases the double periodic nature of the curve. Observe that there is no squircle centered at the origin anymore. Instead there is a nearby squircle for each of the 4 quadrants surrounding the origin.

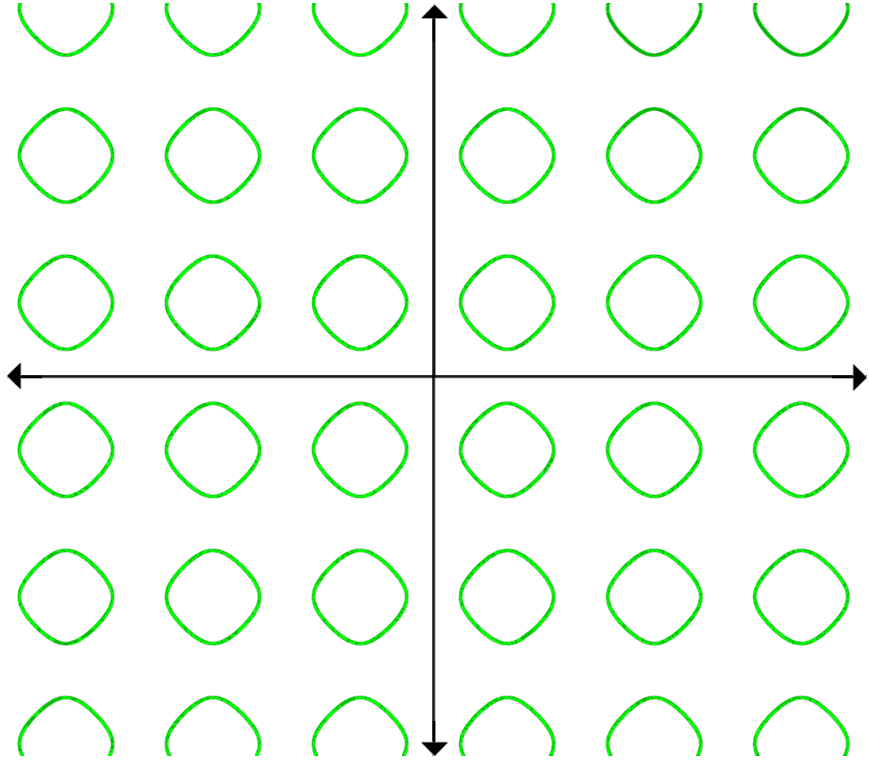

Figure 22: Overshoot introduces a phase shift to the oblique squircle

Just as with the oblique squircle, its 3D counterpart has an analogous overshoot. The amended equation for the implicit surface with overshoot is

$$\cos\left(\frac{s\pi x}{r}\right) + \cos\left(\frac{s\pi y}{r}\right) + \cos\left(\frac{s\pi z}{r}\right) = 2 + \cos(s\pi) - \lfloor s \rfloor h$$

This time, the overshoot variable has a span where $h \in [0,4]$. This occurs because the bounds for the left hand side of the implicit equation

$$-3 \leq \cos\left(\frac{s\pi x}{r}\right) + \cos\left(\frac{s\pi y}{r}\right) + \cos\left(\frac{s\pi z}{r}\right) \leq 3$$

does not match the bounds for the right hand side

$$1 \leq 2 + \cos(s\pi) \leq 3$$

Figure 23 shows surfaces that arise at increasing values of overshoot. Note that all these surfaces appearing in Figure 23 are triply periodic in space, but we restricted them to a unit cell in order to simplify the visualizations. Observe that when $h = 1$, the resulting surface resembles the Schwarz P minimal surface. This follows because for this specific surface, we have squareness $s = 1$, size $r = \pi$, and overshoot $h = 1$. After simplifying, this reduces to the equation

$$\cos x + \cos y + \cos z = 0$$

This implicit equation is a well-known approximation of the Schwarz P minimal surface [Hoffman 1998] [Ferreol 2017a]. We shall refer to it as the *sham Schwarz surface*.

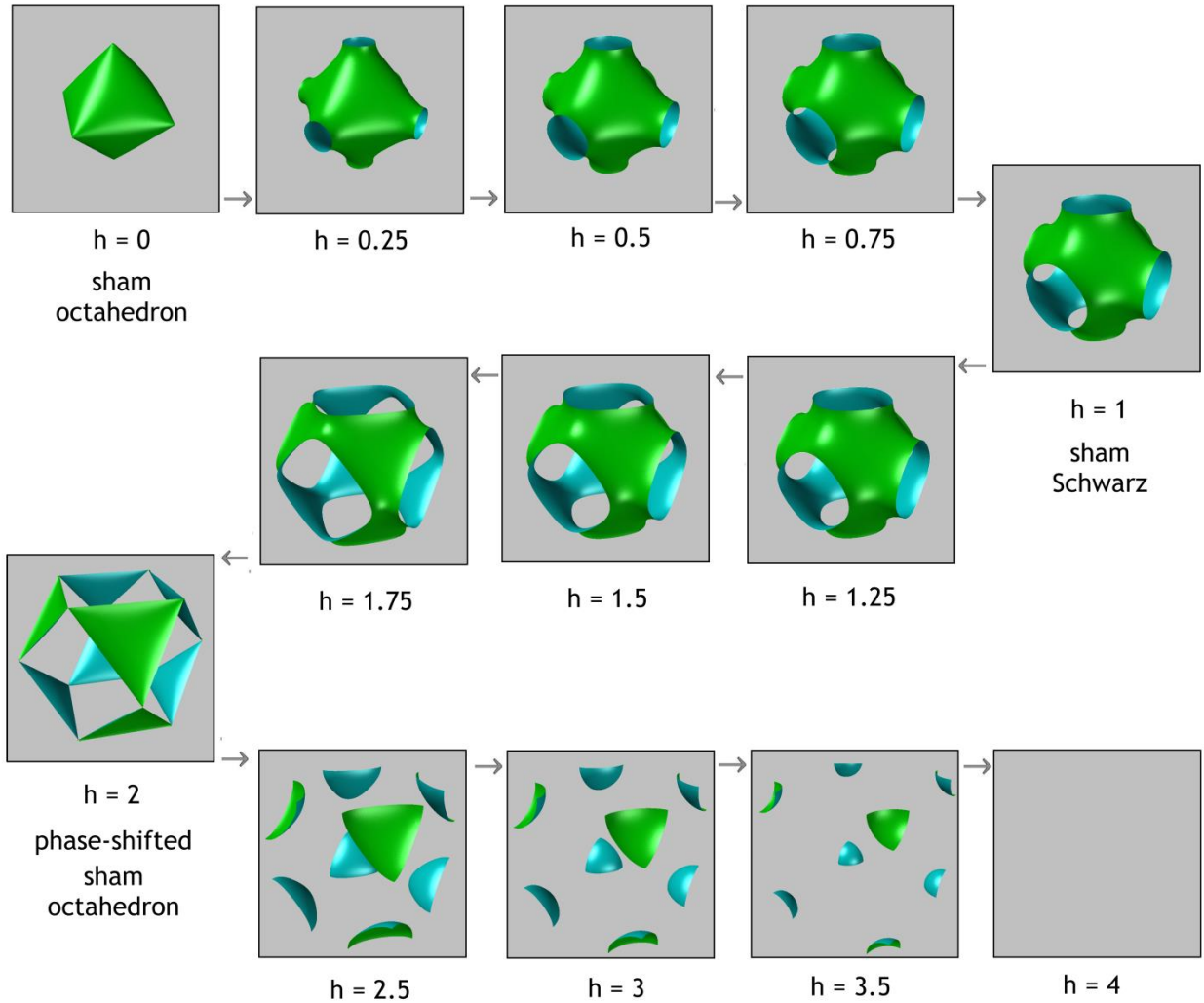

Figure 23: 3D counterpart of the oblique squircle with overshoot

Overshoot in 3D introduces phenomena that are analogous to the 2D case. When $h = 2$, there is a phase shift of the sham octahedron. When $h = 4$, the 3D shape complete recedes into nothing. These phenomena are shown in Figure 23 and can be compared to the analogous 2D case in Figure 21.

In the next section, we will delve more into the Sham Schwarz surface and its relation to the Schwarz P surface. Both surfaces are triply-periodic with octahedral symmetry for their unit cell. Furthermore, we will show that both surfaces are related to the squircle.

# 11. The Schwarz P Surface

In differential geometry, every point of a smooth surface is known to have two principal curvatures $k_1$ and $k_2$. The mean curvature of a point on a smooth surface is defined as $H = \frac{1}{2}(k_1 + k_2)$. Meanwhile, a minimal surface is defined as a surface with zero mean curvature everywhere. The Schwarz P surface is an example of a minimal surface and is shown on the left side of Figure 24. The P in the name of the surface is a shortening of the word "primitive".

We have already encountered triply-periodic surfaces previously when we discussed the 3D counterparts of the periodic squircle and the oblique squircle. The Schwarz P surface is also a triply-periodic surface. It was originally studied by Hermann Schwarz in the 1860s within the context of minimal surfaces. It has the shape of a soap film that arises when a certain skew hexagonal wire frame [Ferreol 2017a] is dipped into liquid soap. In order to calculate points on a minimal surface, one has to solve a partial differential equation (PDE) that minimizes surface area for a given boundary condition. The specific PDE for this was discovered by Joseph-Louis Lagrange in 1762 and has this equation:

$$(1 + z_x^2)z_{yy} - 2z_x z_y z_{xy} + (1 + z_y^2)z_{xx} = 0$$

More recently, researchers in structural chemistry [Gandy 2000] have discovered closed-form analytical expressions for a parametric representation of the Schwarz P surface, but the equations are complicated and involve elliptic integrals with complex parameters. For visualization purposes, it is usually simpler to use an approximation of the Schwarz P surface in the form of an implicit surface with the equation

$$\cos(x) + \cos(y) + \cos(z) = 0$$

As discussed in the previous section, we shall refer to this implicit surface as the *sham Schwarz surface*. . It is a specific 3D counterpart of the oblique squircle with overshoot *h=1*. The sham Schwarz surface is shown on the right side of Figure 24. Note that Schwarz P surface and the sham Schwarz surface are triply-periodic in Euclidean space, but we only show 1 cell of the surface in Figure 24.

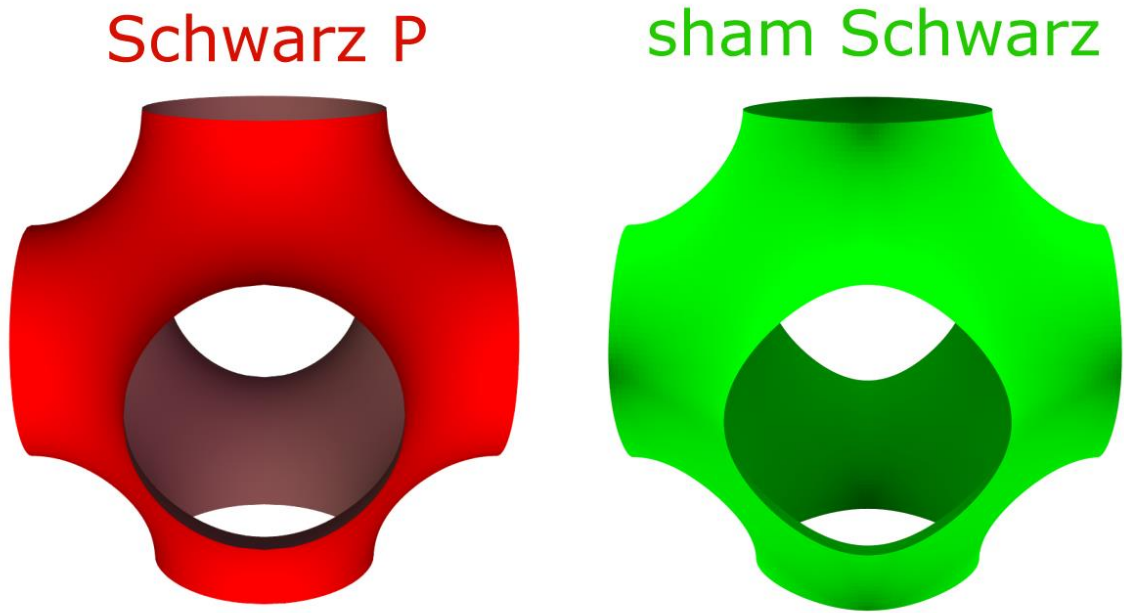

Figure 24: Side-by-side comparison of the Schwarz P surface with the sham Schwarz surface

We are primarily interested in the underlying shape of the 6 holes in both the Schwarz P surface and the sham Schwarz surface. These holes are not really true holes but are actually cross sections where cells interface with other cells in the triply-periodic surfaces. For this reason, we shall refer to these "holes" as *orifices* instead.

When Hermann Schwarz originally studied his namesake primitive surface, he observed that the orifice is almost circular in cross-section. He mentioned that the 2D shape encompassing the orifice had radial errors measuring less than 0.4% from an actual circle [Gandy 2000]. This is quite evident by visual inspection of Figure 24.

In contrast, the orifice of sham Schwarz surface is obviously not circular. Using the implicit equations for the oblique squircle and its 3D counterpart, we can to figure out what shape encompasses its orifice. Recall that the implicit equation for the sham Schwarz surface is $cos(x) + cos(y) + cos(z) = 0$. Also, we know that two of the orifices are located at $z = \pm\pi$. Substituting back to the implicit equation, we get $cos(x) + cos(y) = 1$. This equation is just the oblique squircle with squareness $s = \frac{1}{2}$, size $r = \pi/2$, and no overshoot. Therefore, we can conclude that the shape of the orifice is an oblique squircle! This is illustrated in the leftmost column of Figure 25 where we show a front view, a perspective view, and cross-sectional view of the sham Schwarz surface with orifice.

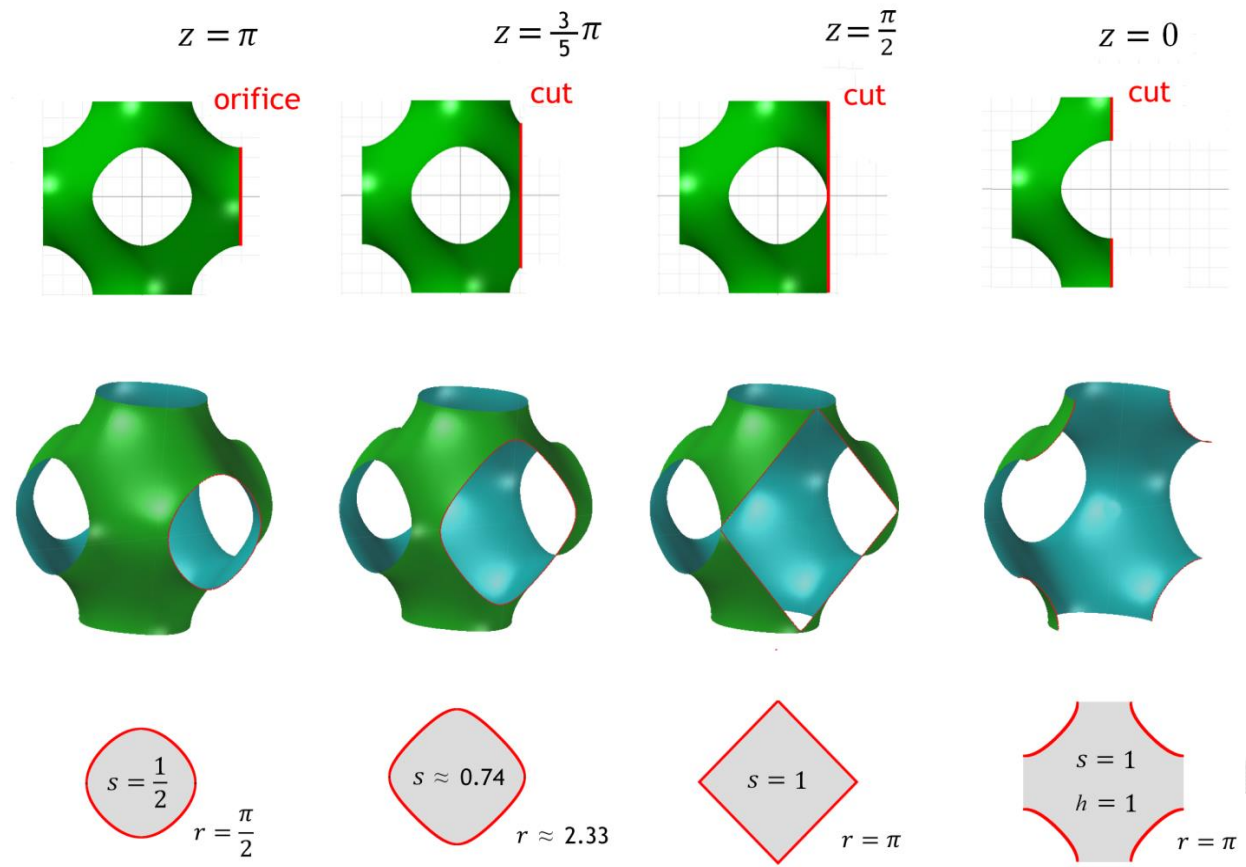

Figure 25: Oblique squircles encompass the perpendicular cross sections of the surface

Using a similar approach, one can deduce that all perpendicular cross sections of the sham Schwarz surface are oblique squircles, with or without overshoot. For example, if we perform a cross-sectional cut at $z = \pi/2$, the implicit equation reduces to the 2D equation: $cos(x) + cos(y) = 0$. This equation is just an oblique squircle with squareness $s = 1$, size $r = \pi$, and no overshoot. The cross section is a square arising from the oblique squircle with a squareness of 1. This is illustrated in the 3rd column of Figure 25.

We have included two other values for cross-sectional cuts in Figure 25. This includes a midway cross-sectional cut at $z = 0$, which produces an oblique squircle with overshoot.

We conclude this section by sharing an observation about the cross sections of the Schwarz P surface and proposing a conjecture. It is known that the Schwarz P surface has embedded lines across the surface [Weber 2013]. These embedded lines can join to form a square cross section, as shown on the left side of Figure 26. Meanwhile, we have observed that if this perpendicular cross section is moved forward with a slight offset, it becomes squircular in shape. This is shown on the right side of Figure 26. We believe that the perpendicular cross sections of the Schwarz P surface form some kind of squircle. This is restated as a conjecture below.

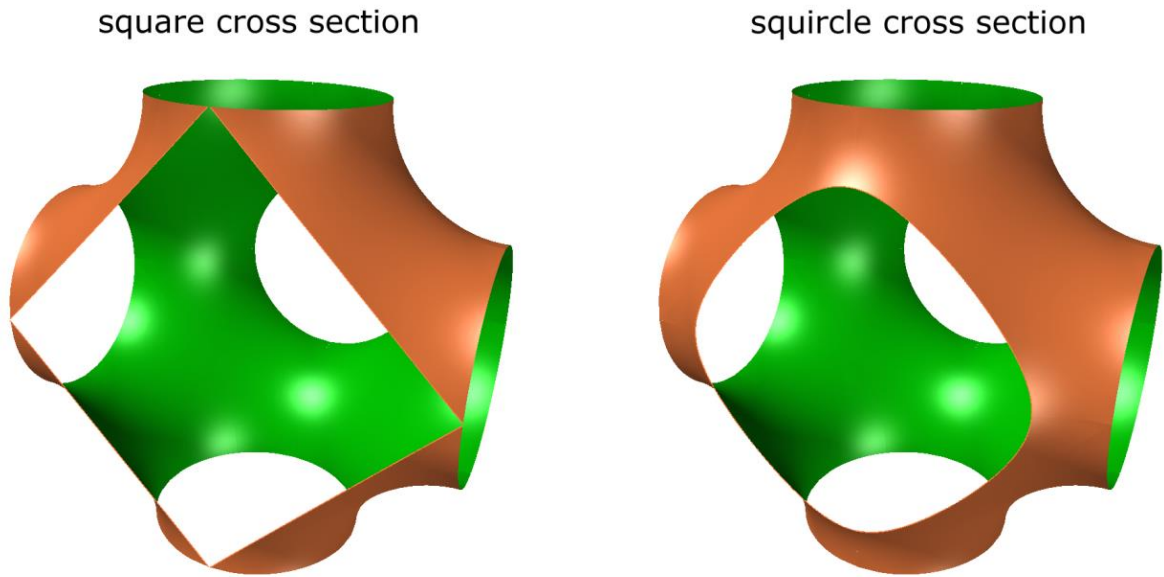

Figure 26: Some perpendicular cross sections of the Schwarz P surface

**Conjecture:** The perpendicular cross-sections of the Schwarz P surface form some kind of squircular shape analogous to the oblique squircle. This occurs, specifically, in the regions where

$$\frac{\pi}{2} \leq |x| \leq \pi$$

$$\frac{\pi}{2} \leq |y| \leq \pi$$

$$\frac{\pi}{2} \leq |z| \leq \pi$$

# 12. Other Squircular Implicit Surfaces

In this section, we will explore more implicit surfaces based on squircles. Similar shapes have been studied before [Barr 1981] [Barr 1992] [Fong 2018], but this paper focuses on using novel squircles discussed earlier as base shapes to come up with new equations for implicit surfaces.

## 12.1 Squircular Toroid

The torus is a common geometric primitive used in many CAD and graphics programs. It is worthwhile to study squircular extensions of it. Consider a torus with a distance of $R$ from the center of its tube to the center of its hole; and $r$ as the radius of its tube. There is a well-established implicit equation for the torus given by:

$$\left(\sqrt{x^2 + y^2} - R\right)^2 + z^2 = r^2$$

This can then be algebraically manipulated to remove the square root operator in the equation and get this equivalent quartic equation:

$$(x^2 + y^2 + z^2 + R^2 - r^2)^2 = 4R^2(x^2 + y^2)$$

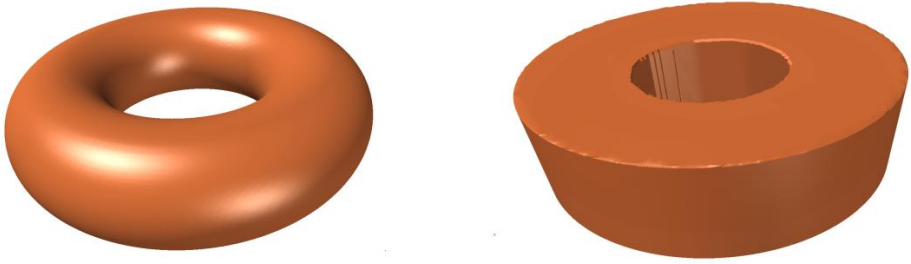

Figure 27: Common toroidal shapes like the torus (left) and the square toroid (right)

Using the Fernandez-Guasti squircle with squareness $s$, once can extend the torus to toroidal shapes with squircular cross sections. For example, a square toroid is shown on the right side of Figure 27. The implicit equation for the squircular toroid is given by

$$\left(\sqrt{x^2 + y^2} - R^2\right)^2 + z^2 - \frac{s^2 z^2}{r^2}\left(\sqrt{x^2 + y^2} - R^2\right)^2 = r^2$$

Just like with the torus, the equation can then be algebraically manipulated to remove the square root operator and get this equivalent octic equation:

$$\left(x^2 + y^2 + z^2 + R^2 - r^2 - \frac{s^2 z^2}{r^2}(x^2 + y^2 + R^2)\right)^2 = 4R^2(x^2 + y^2)\left(1 - \frac{s^2 z^2}{r^2}\right)^2$$

Furthermore, one can vary the squareness parameter in the equation to get different squircular cross sections for the shape. To illustrate this, we show the half-toroid at different squareness values in Figure 28.

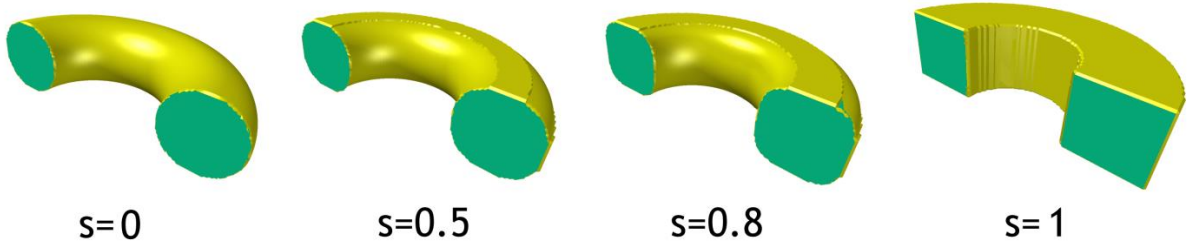

Figure 28: Squircular half-toroid at varying squareness values

## 12.2 Squircular Cone

The Fernandez-Guasti squircle has previously been used to model a cone with a squircular base [Fong 2018], the equation for this is

$$x^2z^2 + y^2z^2 - s^2c^2x^2y^2 - \frac{z^4}{c^2} = 0$$

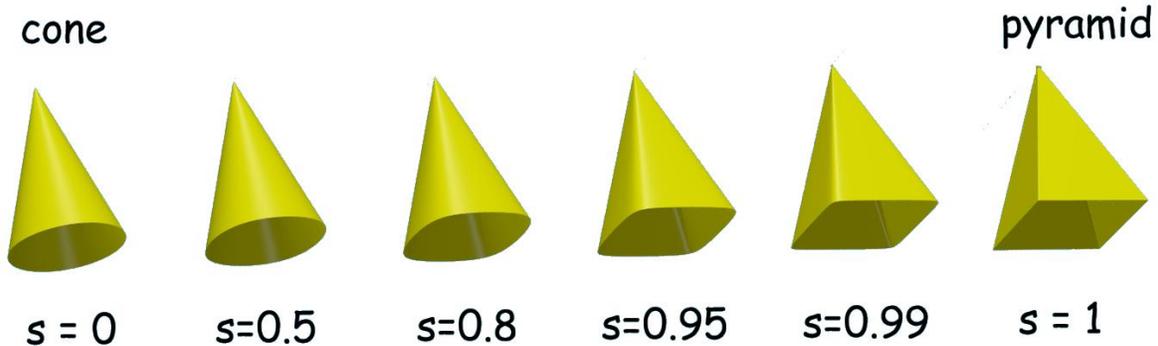

Figure 29: A squircular cone based on Fernandez-Guasti squircle with $c = 3.0$

In addition to the implicit equation provided above, we need to subject the squircular cone to some inequalities that bound x, y, and z in order to prune away extraneous parts in the shape.

$$0 \le z \le c \qquad\qquad -\frac{z}{c} \le x \le \frac{z}{c} \qquad\qquad -\frac{z}{c} \le y \le \frac{z}{c}$$

It is possible to use Lamé lower squircle instead of the Fernandez-Guasti squircle to come-up with an alternative equation for a squircular cone. The implicit equation for this is

$$\left|\frac{x}{a}\right|^p + \left|\frac{y}{b}\right|^p = \left|\frac{z}{c}\right|^p \qquad with \qquad \begin{array}{l} 1 \le p \le 2 \\ 0 \le z \le c \end{array}$$

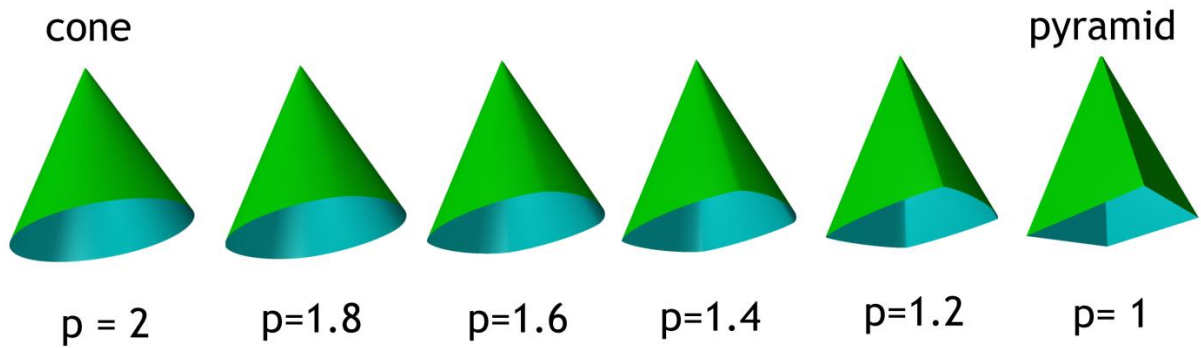

Figure 30: A squircular cone based on the Lamé lower squircle with $a=b=1$ and $c=2$

This shape is shown in Figure 30 with varying $p$ exponents. The constants $a$ and $b$ specify the semi-major and semi-minor lengths of the base shape. When $a = b$, the base shape has no eccentricity. The constant $c$ specifies the height of the cone. When $p = 2$, the 3D shape is a cone. When $p = 1$, the 3D shape is square pyramid.

One advantage of using the Lamé lower squircle for the base shape is that it does not involve infinity in representing the square. This is unlike the Lamé upper squircle which requires infinite limits within its equations.

## 12.3 Sham Cuboctahedron

There has been interest in modeling and approximating various polyhedra using implicit surfaces. For example, the Goursat surface [Ferreol 2017b] has been used to produce rounded approximations of the cube, octahedron, dodecahedron, and icosahedron. Meanwhile, the Cayley surface [Ferreol 2017c] has been used to produce rounded approximations of the regular tetrahedron.

In this subsection, we will introduce an implicit surface that is a rounded approximation of the cuboctahedron. This surface, which we shall name the *sham cuboctahedron*, comes from a modification of the equation for the *sphube*. The sphube was previously discussed as the 3D counterpart of the Fernandez-Guasti squircle. The implicit equation for the sham cuboctahedron is

$$x^2 + y^2 + z^2 - x^2y^2 - y^2z^2 - x^2z^2 + 2x^2y^2z^2 = 1 \qquad with \qquad \begin{array}{l} -1 \leq x \leq 1 \\ -1 \leq y \leq 1 \\ -1 \leq z \leq 1 \end{array}$$

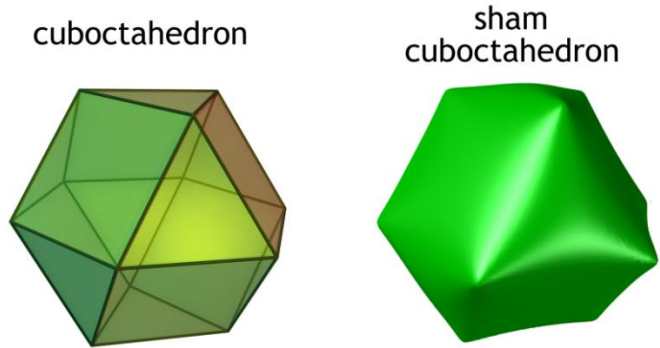

Figure 31: An implicit surface approximation of the cuboctahedron

This sham cuboctahedron is sextic surface. The surface has 12 singularities corresponding to each of the vertices of the cuboctahedron. A side by side comparison with the cuboctahedron is shown in Figure 31. If scale is incorporated into the implicit equation, it becomes

$$\frac{x^2}{k^2} + \frac{y^2}{k^2} + \frac{z^2}{k^2} - \frac{x^2y^2}{k^4} - \frac{y^2z^2}{k^4} - \frac{x^2z^2}{k^4} + \frac{cx^2y^2z^2}{k^6} = 1$$

where *k* is the constant scale factor and *c* is a constant not necessarily equal to 2. Heuristically, we have found that *1.5 ≤ c ≤ 4* gives an acceptable approximation of the cuboctahedron.

## 13. Summary

We introduced two new types of squircles – the periodic squircle and the oblique squircle, along with their 3D counterparts. We also used different types of squircles as base shapes to come up with novel implicit surfaces.